\documentclass[conference,a4paper]{IEEEtran}
\IEEEoverridecommandlockouts
\usepackage{cite}
\usepackage{amsmath,amssymb,amsfonts}
\usepackage{algorithmic}
\usepackage{graphicx}
\usepackage{textcomp}
\usepackage{xcolor}
\usepackage[inline]{enumitem}

\usepackage[top=20mm,
            bottom=44mm,
            left=14mm,
            right=14mm]{geometry}

\def\BibTeX{{\rm B\kern-.05em{\sc i\kern-.025em b}\kern-.08em
    T\kern-.1667em\lower.7ex\hbox{E}\kern-.125emX}}

\begin{document}

\title{Contextualized AI for Cyber Defense:\\An Automated Survey using LLMs}

\author{
    \IEEEauthorblockN{Christoforus Yoga Haryanto}
    \IEEEauthorblockA{\textit{School of Science} \\
    \textit{RMIT University}\\
    Melbourne, Australia \\
    0009-0009-8340-5313}
    \\
    \IEEEauthorblockN{Yoshiano Hartanto}
    \IEEEauthorblockA{\textit{Faculty of Engineering and IT} \\
    \textit{University of Tehnology}\\
    Sydney, Australia \\
    }
    \and
    \IEEEauthorblockN{Anne Maria Elvira}
    \IEEEauthorblockA{\textit{School of Science} \\
    \textit{RMIT University}\\
    Melbourne, Australia \\ \\
    }
    \\
    \IEEEauthorblockN{Emily Lomempow}
    \IEEEauthorblockA{
    \textit{ZipThought}\\
    Melbourne, Australia \\
    emily@zipthought.com.au}
    \and
    \IEEEauthorblockN{Trung Duc Nguyen}
    \IEEEauthorblockA{\textit{School of Science} \\
    \textit{RMIT University}\\
    Melbourne, Australia \\ \\
    }
    \\
    \IEEEauthorblockN{Arathi Arakala}
    \IEEEauthorblockA{\textit{School of Science} \\
    \textit{RMIT University}\\
    Melbourne, Australia \\
    arathi.arakala@rmit.edu.au}
    \and
    \IEEEauthorblockN{Minh Hieu Vu}
    \IEEEauthorblockA{\textit{School of Science} \\
    \textit{RMIT University}\\
    Melbourne, Australia \\
    }
}

\maketitle

\begin{abstract}
This paper surveys the potential of contextualized AI in enhancing cyber defense capabilities, revealing significant research growth from 2015 to 2024. We identify a focus on robustness, reliability, and integration methods, while noting gaps in organizational trust and governance frameworks. Our study employs two LLM-assisted literature survey methodologies: (A) ChatGPT 4 for exploration, and (B) Gemma 2:9b for filtering with Claude 3.5 Sonnet for full-text analysis. We discuss the effectiveness and challenges of using LLMs in academic research, providing insights for future researchers.
\end{abstract}


\section{Introduction}
Contextualized AI enhances traditional Artificial Intelligence (AI) and Large Language Model (LLM) capabilities by integrating private data, beyond typical public datasets \cite{Pinto2024}. This emerging field finds application in autonomous monitoring, threat detection, and response within secured network environments \cite{Ahmad2023}, \cite{Pearce2023}, \cite{Sasaki2023}, \cite{Loevenich2024}. Yet, the full efficacy of these systems is under ongoing evaluation with challenges such as AI dependency, data privacy, human oversight, and end-to-end governance \cite{Bubeck2023}, \cite{Carlini2020}, \cite{Ganguli2022}, \cite{Gupta2023}, \cite{Oh2023}, \cite{Sanderson2023}, \cite{Haryanto2024a}.

Initially, we hypothesized such systems were ready for widespread deployment in cyber security. However, we discovered a complex landscape with diverse terminology and approaches, leading us to do comprehensive literature review. Given the vast amount of potentially relevant research and the challenges in identifying pertinent studies, we decided to experiment with LLM-assisted methods for our survey while also providing us an opportunity to explore innovative methodologies for academic research \cite{Agarwal2024}, \cite{Joos2024}, \cite{Haryanto2024b}.

This paper aims to conduct a survey of the literature in contextualized AI for cyber defense to answer:
\begin{enumerate}
    \item RQ1: How can cyber security decision-makers strategically leverage contextualized AI to enhance defense capabilities while mitigating risks?
    \item RQ2: Protection Layer: How can we ensure comprehensive protection of the system, data, and processes when implementing contextualized AI in cyber security?
    \item RQ3: Security System Layer: How can we guarantee that the AI-enhanced protection system itself functions reliably and as expected?
    \item RQ4: Organizational Layer: How can we foster organizational trust in AI, ensuring that the organization can confidently rely on AI capabilities within appropriate scopes while maintaining necessary human oversight?
\end{enumerate}

\subsection{Definition of Contextualized AI}
For this study, we define contextualized AI to be supplied as part of the prompt to LLM as follows:

\small
\begin{verbatim}
Contextualized AI refers to AI systems 
designed to access and utilize proprietary 
and domain-specific knowledge. While it is 
not strictly generative AI and LLM, most of
the contextualized AI systems are built on 
top of generative AI and LLM so pay attention
to the usage that involves further training
using proprietary or domain-specific knowledge
on top of pre-trained model. Some earlier
papers may mention full AI training using
private data, hence they should be considered
as contextualized AI systems too.
\end{verbatim}
\normalsize

\subsection{Paper Structure}
Section II describes our method, including the use of LLM tools, Section III discusses the findings from exploration using GPT-4, Section IV and V discusses the findings from literature screening using Gemma 2:9b and full-text analysis using Claude 3.5 Sonnet, and Section VI summarizes all the findings, including an analysis to the research methodology we use, and recommends for the future research directions.

This paper presents two distinct methodologies for LLM-assisted literature surveys: Method \texttt{A} using ChatGPT 4 for initial exploration and thematic analysis, and Method \texttt{B} combining Gemma 2:9b for literature screening with Claude 3.5 Sonnet for full-text analysis. We compare these approaches to demonstrate their effectiveness in processing large volumes of academic literature efficiently.

\section{Methodologies}
\begin{figure}[!b]
\centering
\includegraphics[width=8cm]{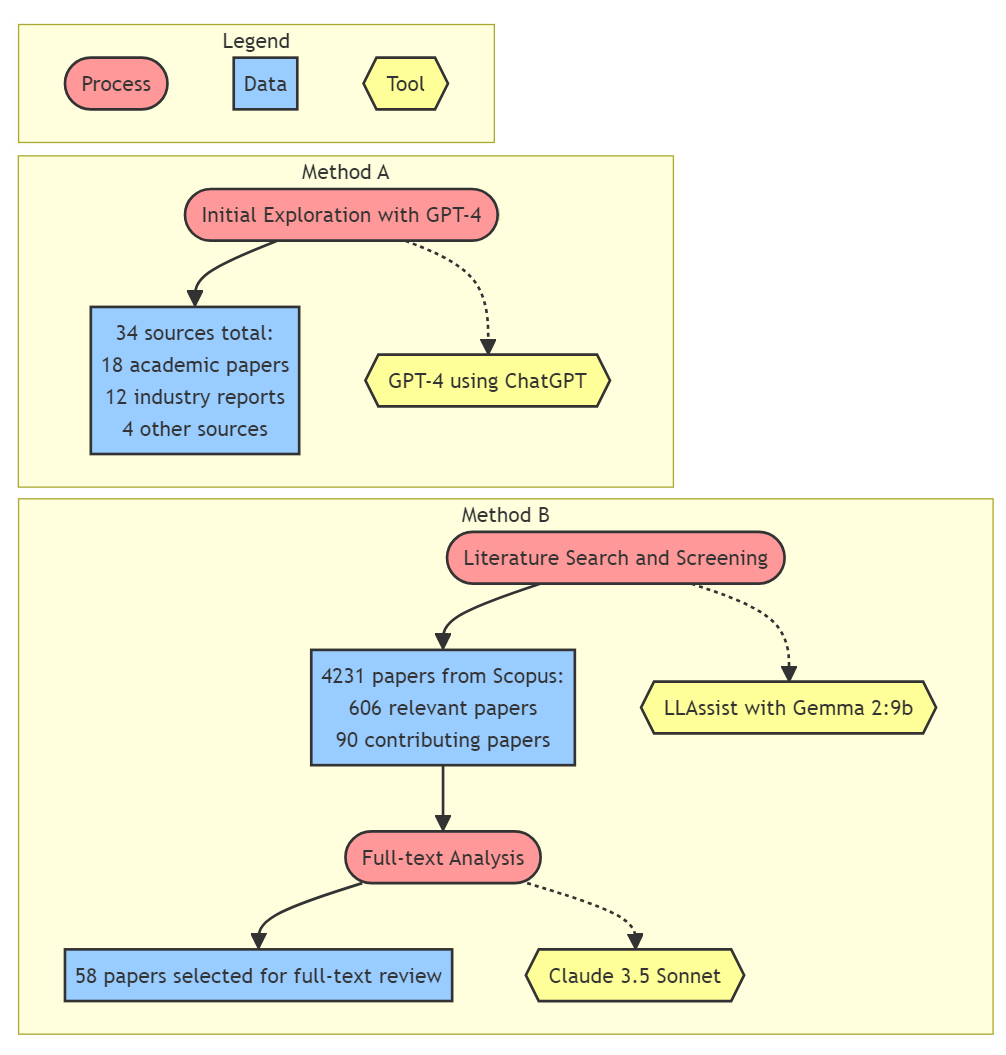}
\caption{Methodology and Data Gathering}
\label{fig:methodology}
\end{figure}

We employ two distinct LLM-assisted approaches:
\begin{enumerate*}
    \item Method A: GPT-4 for initial exploration and thematic analysis.
    \item Method B: Gemma 2:9b for literature screening and Claude 3.5 Sonnet for full-text analysis.
\end{enumerate*}
See Fig. \ref{fig:methodology} for the overview. 

\subsection{Method A: Exploration with GPT}
Method \texttt{A} uses GPT-4 via ChatGPT to rapidly generate a broad overview and identify key themes using the following steps:

\begin{enumerate}
    \item Input research questions and definitions into GPT-4.
    \item Instructed GPT-4 to search for relevant literature.
    \item Used consistent prompt structure for each of RQ:
    \small \begin{verbatim}
{{Research Question}}

{{Definitions}}

Search online for relevant conference
papers and journal articles.
    \end{verbatim}\normalsize
    \item Prompted twice more with \texttt{Find more} for each query.
    \item Collected and categorized provided sources.
\end{enumerate}

Our exploratory review with GPT-4 returned 34 unique sources without dead links, including 18 academic publications (52.9\%), 12 industry reports (35.3\%), 2 professional organization resources (5.9\%), and 2 non-profit think tank publications (5.9\%). All the returned sources were publicly accessible.

\subsection{Method B: Systematic Review with Gemma and Claude}

Method \texttt{B} employs a more systematic approach Gemma 2:9b and Claude 3.5 Sonnet for in-depth analysis \cite{Haryanto2024b}, \cite{gemma2}, \cite{claude35sonnet}. We chose Gemma 2:9b for literature screening due to its efficiency in processing large volumes of text. Claude 3.5 Sonnet was selected for its ability to do full-text analysis of academic papers using prompt engineering.

\textbf{Literature Screening with Gemma 2:9b:} We used LLAssist, an LLM-based simplified screening tool with Gemma 2:9b backend (commit version \texttt{3bf51a6}) \cite{llassist}. LLAssist streamlines literature reviews through the following process:

\begin{enumerate}
    \item \textbf{Data Input:} Processes CSV files with article metadata and abstracts, along with research questions.
    \item \textbf{Key Semantics Extraction:} Extracts topics, entities, and keywords from titles and abstracts using Natural Language Processing (NLP).
    \item \textbf{Relevance Estimation:} Assesses each article's relevance to research questions, providing scores (0-1) for relevance and contribution, with 0.7 as the threshold between false (below 0.7) and true (0.7 and above).
    \item \textbf{Must-Read Determination:} Identifies "must-read" articles based on relevance and contribution scores.
    \item \textbf{Output Generation:} Produces JSON and CSV files with detailed information for each article.
\end{enumerate}

The process involved:

\begin{enumerate}
    \item Querying the Scopus database with search string as \texttt{"artificial AND intelligence AND cyber AND security"} for 2015-2024. We use Scopus as its result already includes IEEE and ACM databases.
    \item To check if the abstract addresses one or more research question we developed screening questions (SQ) for the LLM prompt as below, followed by the definitions:
    \begin{itemize}
        \item SQ1: Does it discuss strategic factors for implementing LLM-based or contextualized AI in cyber security defense? [\texttt{{Definitions}}]
        \item SQ2: Does it mention methods for integrating such AI into cyber security defense systems? [\texttt{{Definitions}}]
        \item SQ3: Does it address techniques for ensuring robustness and reliability of these AI systems? [\texttt{{Definitions}}]
        \item SQ4: Does it discuss organizational measures or governance frameworks for building trust in these solutions? [\texttt{{Definitions}}]
    \end{itemize}
    \item Analyze relevance and contribution of papers based on Gemma 2:9b's scores, using 0.7 as the threshold.
\end{enumerate}

\textbf{In-depth Analysis with Claude 3.5 Sonnet:} We conducted a full-text review using Claude 3.5 Sonnet \cite{claude35sonnet} with prompting techniques \cite{Askell2021}. The following prompt engineering techniques are used: "Prompt generator for the initial draft", "Be clear and direct", "Give Claude a role", "Prefill Claude's response", and "Long context prompting" \cite{claudeprompteng}. We use the following steps:

\begin{enumerate}
    \item Filtering contributing papers from 2020-2024 with DOIs, excluding non-Q1 journal articles.
    \item Analyzing full-text PDFs using Claude 3.5 Sonnet.
    \item Manual review for divergences between AI and human interpretations. We read the paper title, abstract, introduction, and conclusion then compare with the LLM output to check whether it is reasonably correct.
\end{enumerate}

We used the following instruction prompt to guide Claude:

\small
\begin{quote}
\begin{verbatim}
You are the research assistant. Given the
research article, explain concisely how
the paper addresses the following research
questions:

{{Research Questions}}

{{Definition of Contextualized AI}}

In your explanation, first read the title
and abstract, determine the type of the
paper, e.g. survey paper, and then read
the introduction and conclusion sections.
Indicate if you need also to read the
entire content of the paper. For each
point you make, give the exact reference
on where the original statement can be
found, i.e. page/section number and
paragraph. Strictly limit it to the actual
content of the paper itself. Put a note in
your explanation if the paper does not
address one or more RQ.

Your output will be:

{{Output Format}}
\end{verbatim}
\end{quote}
\normalsize

This prompt engineering approach is effective in extracting relevant information and insights from full-text articles.

\section{Exploration with GPT-4}
\subsection{Key Findings}
Our exploration with GPT-4 yielded valuable insights for each research question, drawing from both academic and non-academic sources:

\subsubsection{RQ1: Strategic Leverage of Contextualized AI}
Academic research highlights enhanced threat detection and adaptive defense \cite{s3_springer1,s3_springer2}, automation of routine tasks \cite{s3_springer1}, and proactive threat prediction \cite{s3_springer2}. Industry perspectives emphasize domain-specific models for improved threat recognition \cite{s3_ebiquity}, human-AI teaming for optimal decision-making \cite{s3_ar5iv1}, and the importance of ethical considerations \cite{s3_springer1,s3_springer2,s3_weforum}.

\subsubsection{RQ2: Comprehensive Protection Strategies}
Academic sources advocate for deep learning in behavioral analysis \cite{s3_springer3,s3_springer4} and integration of generative AI technologies \cite{s3_ar5iv2}. Industry reports recommend persistent monitoring and real-time analysis \cite{s3_sophos,s3_crowdstrike}, contextualized security measures \cite{s3_wipro}, AI-driven anomaly detection \cite{s3_sophos}, and automation of security processes \cite{s3_wipro,s3_secureframe}.

\subsubsection{RQ3: Ensuring AI System Reliability}
Academic research focuses on robust system structures and reliability assessment \cite{s3_ar5iv3}, recurrent events analysis for prediction \cite{s3_oxford1}, and Scientific Machine Learning (SciML) for safeguarding \cite{s3_oxford1}. Industry and government initiatives emphasize enhanced protection strategies like GREP \cite{s3_researchr}, AI Systems Engineering and Reliability Technologies (ASERT) \cite{s3_llmit}, risk management \cite{s3_createprogress}, and human-in-the-loop approaches with continuous monitoring \cite{s3_csa}.

\subsubsection{RQ4: Fostering Organizational Trust}
Academic sources stress transparency and explainability of AI systems \cite{s3_mdpi,s3_springer5}, ethical considerations in AI development \cite{s3_springer6,s3_springer7}, and dynamic trust calibration mechanisms \cite{s3_ar5iv4}. Industry perspectives highlight robust compliance and security implementation \cite{s3_sumoanalytics}, cultivating a culture of ethical AI use \cite{s3_nextgeninvent}, effective communication strategies \cite{s3_ar5iv5}, talent development \cite{s3_deloitte1,s3_deloitte2}, and stakeholder engagement with policy advocacy \cite{s3_rand,s3_weforum}.

\subsection{Common Themes}
While academic and non-academic sources address similar themes, the angles are different. Academic sources tend to focus on theoretical frameworks, in-depth technical aspects, and long-term implications while non-academic sources emphasize practical applications with market-orientation. Across both academic and industry sources, common themes emerged:
\begin{itemize}
    \item Importance of human-AI collaboration \cite{s3_ar5iv1}, \cite{s3_newsroom}, \cite{s3_cutter}
    \item Continuous learning and adaptation \cite{s3_springer2}, \cite{s3_csa}, \cite{s3_deloitte3}
    \item Ethical considerations and transparency \cite{s3_springer1}, \cite{s3_weforum}, \cite{s3_mdpi}
    \item Balancing automation and oversight \cite{s3_wipro}, \cite{s3_newsroom} \cite{s3_vectra}
    \item Significance of contextual understanding in AI \cite{s3_ebiquity}, \cite{s3_springer3}
\end{itemize}

\section{Analysis of Gemma2:9b Filtering Result}
\begin{figure}[!b]
\centering
\includegraphics[width=8cm]{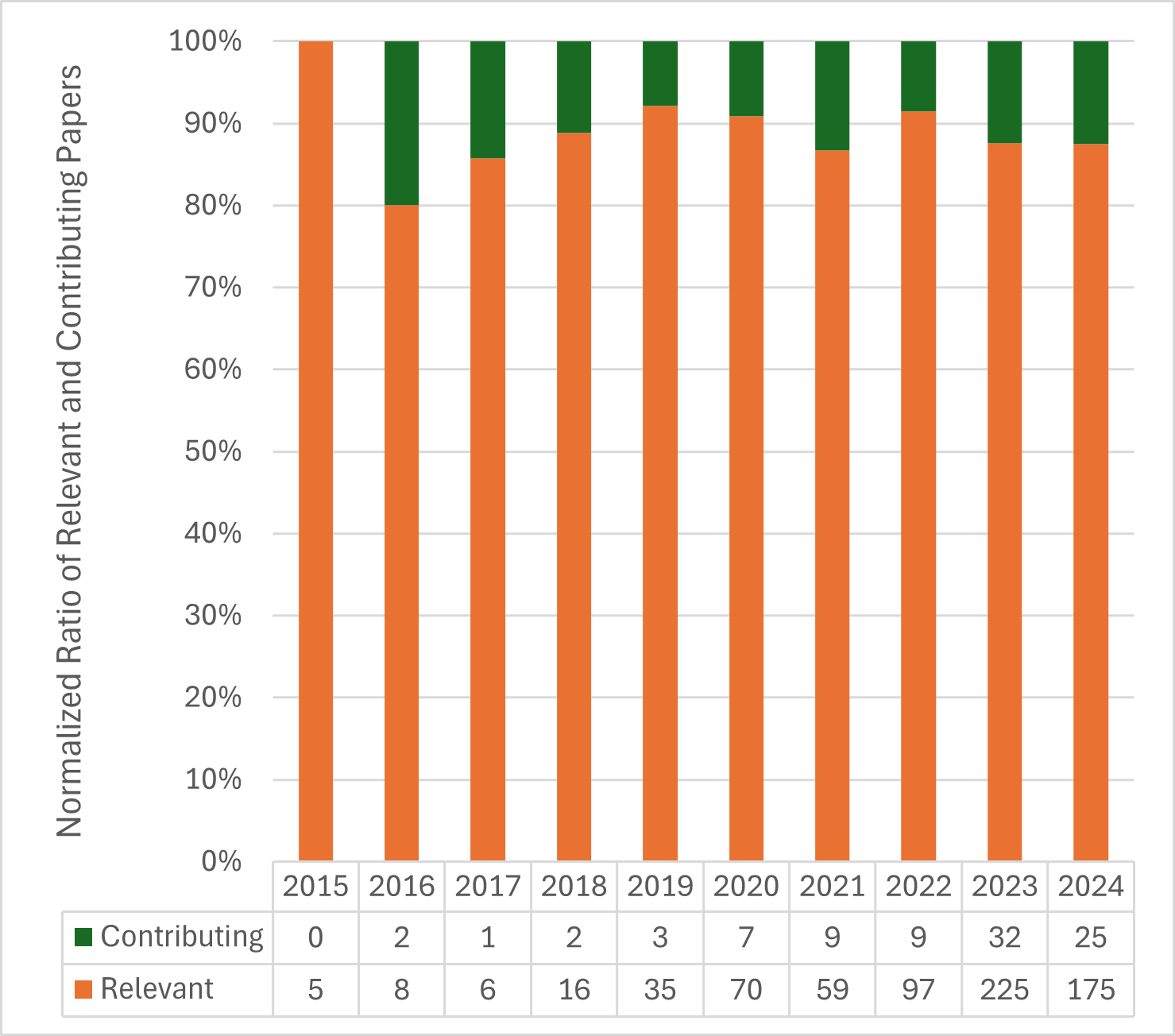}
\caption{Ratio between Relevant and Contributing Papers}
\label{fig:distribution_RC}
\end{figure}

Our analysis, based on LLAssist's output applied to Scopus database results, uses a scoring system (0-1) for relevance and contribution, with 0.7 as the threshold for classification. Tables \ref{tab:data_summary_R} and \ref{tab:data_summary_C} summarize relevant and contributing papers from 2015-2024, respectively. Figure \ref{fig:distribution_RC} visualizes the distribution of both categories. In the tables, we define:
\begin{enumerate*}
    \item \texttt{SQ}: Screening Question
    \item \texttt{R}: Relevant papers, i.e. papers discussing topics related to the screening questions
    \item \texttt{C}: Contributing papers, i.e. papers directly researching topics in the screening questions
    \item \texttt{Any SQ}: Papers relevant to or contributing to at least one screening question
    \item \texttt{All SQs}: Papers relevant to or contributing to all screening questions
\end{enumerate*}

The relevance criterion assesses whether a paper discusses topics related to our research questions, while the contribution criterion evaluates whether a paper directly researches these topics. We use a score threshold of 0.7 for both criteria to determine if a paper is considered relevant or contributing. The analysis reveals significant growth in research attention, with relevant papers increasing from 5 (2015) to 150 (2024), and contributing papers from 0 to 25 over the same period.

\begin{table}[!htbp]
\centering
\caption{Distribution of Relevant Papers (R) by Year and SQ}
\label{tab:data_summary_R}
\begin{tabular}{|r|r|r|r|r|r|r|r|}
\hline
\textbf{Year} & \textbf{Total} & \textbf{Any SQ} & \textbf{All SQs} & \textbf{SQ1} & \textbf{SQ2} & \textbf{SQ3} & \textbf{SQ4} \\
\hline
2015 & 81 & 5 & 0 & 2 & 2 & 3 & 0 \\
2016 & 125 & 6 & 0 & 1 & 2 & 5 & 2 \\
2017 & 167 & 5 & 1 & 3 & 1 & 4 & 1 \\
2018 & 255 & 14 & 2 & 9 & 9 & 8 & 3 \\
2019 & 315 & 32 & 1 & 19 & 10 & 21 & 3 \\
2020 & 380 & 63 & 3 & 31 & 30 & 46 & 13 \\
2021 & 485 & 50 & 9 & 31 & 29 & 36 & 12 \\
2022 & 720 & 88 & 5 & 40 & 51 & 61 & 12 \\
2023 & 1078 & 193 & 18 & 92 & 116 & 105 & 41 \\
2024 & 625 & 150 & 27 & 87 & 102 & 87 & 44 \\
\hline
Total & 4231 & 606 & 66 & 315 & 352 & 376 & 131 \\
\hline
\end{tabular}
\end{table}

\begin{table}[!t]
\centering
\caption{Distribution of Contributing Papers (C) by Year and SQ}
\label{tab:data_summary_C}
\begin{tabular}{|r|r|r|r|r|r|r|r|}
\hline
\textbf{Year} & \textbf{Total} & \textbf{Any SQ} & \textbf{All SQs} & \textbf{SQ1} & \textbf{SQ2} & \textbf{SQ3} & \textbf{SQ4} \\
\hline
2015 & 81 & 0 & 0 & 0 & 0 & 0 & 0 \\
2016 & 125 & 2 & 0 & 1 & 0 & 1 & 0 \\
2017 & 167 & 1 & 0 & 0 & 0 & 1 & 0 \\
2018 & 255 & 2 & 0 & 1 & 0 & 1 & 0 \\
2019 & 315 & 3 & 0 & 3 & 0 & 1 & 0 \\
2020 & 380 & 7 & 0 & 4 & 0 & 3 & 0 \\
2021 & 485 & 9 & 0 & 6 & 1 & 8 & 0 \\
2022 & 720 & 9 & 0 & 6 & 0 & 3 & 0 \\
2023 & 1078 & 32 & 0 & 15 & 3 & 19 & 1 \\
2024 & 625 & 25 & 0 & 16 & 2 & 11 & 3 \\
\hline
Total & 4231 & 90 & 0 & 52 & 6 & 48 & 4 \\
\hline
\end{tabular}
\end{table}

\subsection{Focus of Research}
Papers on robustness and reliability received the most attention (376 relevant, 48 contributing), followed by integration methods and strategic implementation factors. Organizational measures and governance frameworks for trust-building received the least attention (131 relevant, 4 contributing).

\subsection{Recent Developments}
2023-2024 saw accelerated research activity across all screening questions. 2023 produced 193 relevant and 32 contributing papers, while 2024 already shows 150 relevant and 25 contributing papers, indicating rapidly growing interest.

\subsection{Research Gaps}
SQ4 in Tables \ref{tab:data_summary_R} and \ref{tab:data_summary_C} reveals a significant gap in research on organizational measures and governance frameworks, suggesting opportunities for future research on trust-building aspects of AI adoption in cyber security.

\subsection{Field Maturity}
The ratio of contributing papers to relevant papers (14.9\%) gives insight into the field's maturity. While there is growing interest, as evidenced by the increase in relevant papers, the slower emergence of contributing papers suggests that it is still developing with considerable potential for in-depth studies.

\subsection{Evolution of Research Focus}
2015-2017 saw minimal relevant or contributing papers. 2018-2021 showed gradual increase, especially in strategic factors, integration methods, and robustness techniques. 2022-2024 demonstrated significant attention growth across all questions, particularly in robustness and reliability techniques.

\subsection{Summary of the Analysis}
This analysis highlights rapid growth and evolution in the research. While attention increases across all aspects, more focused research on organizational measures and governance frameworks for trust-building is needed. Recent surges suggest accelerated development with the potential for significant future advancements. For deeper insights into recent impactful research, we conducted a full-text review of selected papers, presented in the following chapter.

\section{Full-text Analysis using Claude 3.5 Sonnet}
Based on the selection criteria, we have short-listed and obtained the full text of 58 research papers between 2020 and 2024. From processing each of the research papers using Claude 3.5 Sonnet, we extracted their type, key themes, author stances, and concise summary of their research results to the RQs and definitions provided as part of the prompt. The key insights are summarized in table \ref{tab:key_insights} and future research direction is outlined in table \ref{tab:future_research}.

\begin{table}[!htbp]
\centering
\caption{Key Insights for Research Questions}
\label{tab:key_insights}
\begin{tabular}{|p{0.7\linewidth}|p{0.2\linewidth}|}
\hline
\textbf{Key Insights} & \textbf{References} \\
\hline
\multicolumn{2}{|l|}{\textbf{RQ1: Key strategic factors to consider}} \\
\hline
AI can significantly enhance system but with new risks & \cite{Gupta2023}, \cite{Orner2024}, \cite{Bécue2021} \\
\hline
Data quality and availability for implementation & \cite{Timofte2024}, \cite{Cody2023} \\
\hline
Interoperability with existing security infrastructure & \cite{Eng2024} \\
\hline
Carefully evaluate ethical and legal implications & \cite{Lee2022}, \cite{Eng2024} \\
\hline
Cost-benefit analysis is essential for decision-making & \cite{Eng2024} \\
\hline
Human oversight and expertise remains critical & \cite{Whyte2022}, \cite{Whyte2023} \\
\hline
\multicolumn{2}{|l|}{\textbf{RQ2: Key integration approaches while preventing overreliance}} \\
\hline
Using AI for automated threat detection and response & \cite{Gupta2023}, \cite{AlHawawreh2023}, \cite{Sai2024} \\
\hline
AI-driven anomaly detection enhances security & \cite{Benzaid2020}, \cite{Jaafar2023} \\
\hline
AI can assist in vulnerability assessment and patching & \cite{Sangalay2024}, \cite{Gasiba2023} \\
\hline
Cyber deception techniques can be enhanced with AI & \cite{Antunnes2023}, \cite{Eng2024} \\
\hline
Human oversight is important to prevent overreliance & \cite{Whyte2022}, \cite{Whyte2023}, \cite{Baroni2021} \\
\hline
Explainable AI improve trust and understanding & \cite{Nadeem2023}, \cite{Zagalsky2021}, \cite{Nguyen2022} \\
\hline
Hybrid of AI and traditional methods are effective & \cite{Tsareva2022}, \cite{Piplai2023} \\
\hline
\multicolumn{2}{|l|}{\textbf{RQ3: Key methods and practices for AI system robustness}} \\
\hline
Adversarial training and testing improve AI robustness & \cite{Kotenko2023}, \cite{Kumar2023}, \cite{Malik2024} \\
\hline
Calibrated uncertainty quantification add reliability & \cite{Fourastier2020}, \cite{Woodward2021} \\
\hline
Deep ensembles and temperature scaling help performance & \cite{Woodward2021} \\
\hline
Need continuous monitoring and adaptation of AI models & \cite{Sample2020}, \cite{Cody2023} \\
\hline
Explainable AI aid in verification and debugging & \cite{Nadeem2023}, \cite{Nguyen2022} \\
\hline
Multiple AI models and perspectives add robustness & \cite{Baroni2021}, \cite{Zagalsky2021} \\
\hline
\multicolumn{2}{|l|}{\textbf{RQ4: Key organizational measures and governance frameworks}} \\
\hline
Need clear policies for AI use in cyber security & \cite{Charfeddine2024}, \cite{Pasupuleti2023}, \cite{McIntosh2024} \\
\hline
Governance needs cross-sector work and standards & \cite{Timofte2024}, \cite{Lee2022} \\
\hline
Regular security assessments and audits of AI systems & \cite{Eng2024} \\
\hline
Transparency and explainability of AI decisions build trust & \cite{Nadeem2023}, \cite{Zagalsky2021} \\
\hline
Compliance with data protection and privacy regulations & \cite{McIntosh2024}, \cite{Sean2024} \\
\hline
Implement ethical guidelines for AI use in cyber security & \cite{Gupta2023}, \cite{Lee2022} \\
\hline
\end{tabular}
\end{table}

Note that during our manual review of the 58 research papers, we encountered several things where our judgment was misaligned with AI, notably \cite{Blowers2020}, \cite{Ranade2021}, \cite{Song2023}, \cite{Nadeem2023}, \cite{Woodward2021}:
\begin{itemize}
    \item There is one occurrence where we deemed the paper is insightful and can contribute towards the robustness \cite{Blowers2020} yet LLM determined the paper as not sufficiently contributing. After careful inspection, we noted that the paper was from 2020, before the advent of LLM-style deep learning. Hence, LLM's determination is correct.
    \item There are two occasions where we failed to identify future research directions \cite{Ranade2021}, \cite{Song2023}. LLM was correct to determine that fake cyber threat intelligence needs deeper research.
    \item There is one occurrence where we misidentified the necessity of privacy-preserving methods in an explainable AI system \cite{Nadeem2023}, dismissing its importance while LLM was correctly identified it as important for future research.
    \item There is one occurrence where we did not fully understand the key insight of using deep ensembles and temperature scaling \cite{Woodward2021}. Upon investigation, the team did not understand those specific terminologies and LLM was correctly identifying them.
\end{itemize}

\begin{table}[t]
\centering
\caption{Research Gaps and Future Directions}
\label{tab:future_research}
\begin{tabular}{|p{0.7\linewidth}|p{0.2\linewidth}|}
\hline
\textbf{Identified Gaps and Future Research Directions} & \textbf{References} \\
\hline
\multicolumn{2}{|l|}{\textbf{RQ1: Future research for strategic decision making}} \\
\hline
Develop a comprehensive decision-making framework & \cite{Malik2024}, \cite{McIntosh2024} \\
\hline
Investigate the long-term impacts of AI adoption & \cite{Bécue2021}, \cite{Gupta2023} \\
\hline
Interplay of AI and evolving threat landscapes & \cite{Benzaid2020}, \cite{Whyte2020} \\
\hline
\multicolumn{2}{|l|}{\textbf{RQ2: Future research for integration approaches}} \\
\hline
Frameworks for balanced human-AI collaboration & \cite{Zagalsky2021}, \cite{Strickson2023}, \cite{Chowdhury2023} \\
\hline
AI-generated fake cyber threat intelligence & \cite{Ranade2021}, \cite{Song2023} \\
\hline
Explore adaptive AI that evolve with threat landscapes & \cite{Timofte2024}, \cite{Cody2023} \\
\hline
\multicolumn{2}{|l|}{\textbf{RQ3: Future research for AI robustness in cyber security}} \\
\hline
Benchmarks and metrics for AI in cyber security & \cite{Malik2024}, \cite{Fenza2024}, \cite{Kumar2024} \\
\hline
Transfer learning and meta-learning approaches & \cite{Antunnes2023} \\
\hline
Privacy-preserving AI for cyber security applications & \cite{Gupta2023}, \cite{Nadeem2023} \\
\hline
Defending against adversarial attacks on AI models & \cite{Kotenko2023}, \cite{Kuppa2020}, \cite{DeHaan2024} \\
\hline
\multicolumn{2}{|l|}{\textbf{RQ4: Future research for organizational readiness}} \\
\hline
Governance frameworks for AI in cyber security & \cite{McIntosh2024}, \cite{Bokhari2023} \\
\hline
Evaluation and certification of AI-driven solutions & \cite{Fenza2024}, \cite{Kumar2024} \\
\hline
Public trust in AI-enhanceed cyber security measures & \cite{Shoaib2023}, \cite{Gupta2023} \\
\hline
Impact of AI on cyber security workforce training & \cite{AlHawawreh2023}, \cite{Charfeddine2024} \\
\hline
\end{tabular}
\end{table}

\section{Discussions and Recommendations}
\subsection{Synthesis of Findings}
Our review reveals a rapidly evolving landscape, with research attention significantly increasing from 2015 to 2024. Key areas requiring further attention:
\begin{enumerate}
\item \textbf{Research Focus Imbalance:} Substantial research exists on technical aspects (robustness, integration), but a gap persists in studies on organizational measures and governance frameworks for building trust in AI-enhanced cyber security solutions.
\item \textbf{Rapid Advancements and Field Maturity:} Research surge from 2022 to 2024 indicates accelerating developments, particularly in integration methods and robustness techniques. The low ratio of contributing to relevant papers suggests a developing field with potential for further substantive contributions.
\end{enumerate}

\subsection{Methodological Analysis and Reflection}
Our AI-assisted exploration and full-text review approach offers both advantages and challenges. Here's a comparison of methodology \texttt{A} (GPT-4 for exploration) and \texttt{B} (Gemma 2:9b for filtering/searching and Claude 3.5 Sonnet for full-text analysis):
\begin{itemize}
    \item \textbf{Immediacy:} \texttt{A} provides immediate thematic review, while \texttt{B} requires a database and full-text access.
    \item \textbf{Efficiency/Breadth:} \texttt{A} processes diverse sources, \texttt{B} focuses on academic literature only.
    \item \textbf{Structure:} \texttt{A} relies on GPT-4's and its search engine result, \texttt{B} uses a consistent assessment framework that can be designed by the researcher.
    \item \textbf{Information Uncovering:} \texttt{A} offers unique cross-source insights, \texttt{B} excels at detailed full-text extraction.
    \item \textbf{Bias Mitigation:} \texttt{A} has both model and search engine bias risk, \texttt{B} has model and academic database bias risk.
    \item \textbf{Context:} \texttt{A} uses general knowledge and instructions, \texttt{B} requires specific prompts and definitions.
    \item \textbf{Depth:} \texttt{A} may sacrifice depth for breadth, \texttt{B} allows in-depth full-text analysis.
    \item \textbf{False Negatives:} \texttt{A} may miss sources not highly ranked in search engine, \texttt{B} may miss non-matching papers.
    \item \textbf{False Positives:} \texttt{A} may include irrelevant sources due to broad interpretations, \texttt{B} minimizes this through filtration.
    \item \textbf{Academic Rigor:} \texttt{A} includes non-academic sources, \texttt{B} focuses on peer-reviewed literature.
\end{itemize}
Both methodologies complement traditional reviews. Future work should refine these techniques while maintaining rigor, potentially:
\begin{itemize}
    \item Developing sophisticated AI prompting strategies
    \item Using multiple AI models for cross-validation
    \item Establishing AI-human analysis integration protocols
    \item Combining strengths of both methodologies
\end{itemize}

\subsection{Implications and Future Directions}
Table \ref{tab:future_research} outlines specific research gaps and future directions. Our analysis reveals rapid field evolution (2022-2024), the need for interdisciplinary approaches, and opportunities for meta-research on AI-assisted systematic reviews in cybersecurity. Future work should prioritize addressing these gaps while balancing innovation and practical implementation.

\section{Conclusion and Future Works}
This survey examined contextualized AI's potential in reshaping cyber defense strategies using a novel AI-assisted methodology. Key findings include significant research attention growth (2015-2024), focus on robustness, reliability, and integration methods, with gaps in organizational trust and governance studies. Our AI-assisted approach demonstrated efficiency in processing diverse sources, highlighting the potential of such methods in comprehensive literature reviews. Future research should prioritize empirical studies comparing traditional and AI-enhanced systems, exploring adaptive AI for evolving threats, and developing governance frameworks. While contextualized AI promises enhanced cyber defense capabilities, effective implementation requires balancing AI strengths with human oversight and risk mitigation. As this field rapidly evolves, interdisciplinary collaboration among cyber security experts, AI researchers, and policymakers will be crucial in addressing the multifaceted challenges of AI contextualization in cyber defense.

\end{document}